\documentclass[11pt]{article}
\usepackage{arxiv}
\usepackage{bbm}
\usepackage[T1]{fontenc}
\usepackage[utf8]{inputenc}
\usepackage[english]{babel}
\usepackage{amsmath,amssymb,amsfonts,amsthm}
\usepackage{cmap}
\usepackage{microtype}
\usepackage{mathtools}
\usepackage{mathrsfs}
\usepackage{bm}

\usepackage[numbers]{natbib}
\usepackage[colorlinks=true]{hyperref}
\hypersetup{
 citecolor=blue,
 linkcolor=red,
 urlcolor=blue}
\usepackage{graphicx}
\usepackage{tabularx}

\usepackage{url}            
\usepackage{booktabs}       
\usepackage{nicefrac}       
\usepackage{microtype}      
\usepackage[inline,shortlabels]{enumitem}
\usepackage[linesnumbered,ruled,vlined]{algorithm2e}
\SetKwInput{KwData}{Input}
\SetKwInput{KwResult}{Output}
\SetKwInput{KwAssumption}{Assumption}

\theoremstyle{plain}



\newtheorem{theorem}{Theorem}
\newtheorem{proposition}{Proposition}

\newtheorem{lemma}[proposition]{Lemma}

\newtheorem{example}[proposition]{Example}

\newtheorem{assumptioninner}{Assumption}
\newenvironment{assumption}[1]{%
  \renewcommand\theassumptioninner{\textsf{#1}}%
  \assumptioninner
}{\endassumptioninner}

\def\GL{\mathrm{GL}}
\def\zarA{\mathscr{A}}

\newcommand{\VV}{V}
\newcommand{\FF}{\mathbb{F}}

\newcommand{\RR}{\mathbb{R}}
\newcommand{\CC}{\mathbb{C}}
\newcommand{\QQ}{\mathbb{Q}}

\newcommand{\rank}{\mathrm{rank}~}
\newcommand{\sign}{\mathrm{sign}~}

\newcommand{\cV}{\mathcal{V}}

\newcommand{\cW}{\mathcal{W}}

\newcommand{\sing}{\mathrm{sing}}

\newcommand{\crit}{\mathrm{crit}}

\newcommand{\jac}{\mathrm{jac}}

\newcommand{\bx}{\bm{x}}

\newcommand{\by}{\bm{y}}
\newcommand{\bma}{{\eta}}
\newcommand{\ff}{\bm{f}}

\newcommand{\DRL}{\mathrm{grevlex}}

\newcommand{\lm}{{\rm lm}}

\title{Faster One Block Quantifier Elimination for Regular Polynomial
Systems of Equations}
\date{\today}
\author{Huu Phuoc {Le} \\
 Sorbonne Universit\'e, \textsc{CNRS},\\
    Laboratoire d'Informatique de Paris~6, \textsc{LIP6}, \\
    \'Equipe \textsc{PolSys} \\
    F-75252, Paris Cedex 05, France \\
    \texttt{huu-phuoc.le@lip6.fr}
\And
Mohab {Safey El Din} \\
Sorbonne Universit\'e, \textsc{CNRS},\\
Laboratoire d'Informatique de Paris~6, \textsc{LIP6}, \\
\'Equipe \textsc{PolSys} \\
F-75252, Paris Cedex 05, France \\
\texttt{mohab.safey@lip6.fr}
}

\begin{document}
\maketitle
\setlength{\parindent}{20pt}
\begin{abstract}
Quantifier elimination over the reals is a central problem in
computational real algebraic geometry, polynomial system solving and
symbolic computation.  Given a semi-algebraic formula (whose atoms are
polynomial constraints) with quantifiers on some variables, it
consists in computing a logically equivalent formula involving only
unquantified variables. When there is no alternation of quantifiers,
one has a \emph{one block} quantifier elimination problem.

This paper studies a variant of the one block quantifier
elimination in which we compute an almost equivalent formula of the
input. We design a new probabilistic efficient algorithm for solving
this variant when the input is a system of polynomial equations
satisfying some regularity assumptions. When the input is generic,
involves $s$ polynomials of degree bounded by $D$ with $n$ quantified
variables and $t$ unquantified ones, we prove that this algorithm
outputs semi-algebraic formulas of degree bounded by $\mathcal{D}$
using $O\ {\widetilde{~}}\left ((n-s+1)\ 8^{t}\ \mathcal{D}^{3t+2}\
\binom{t+\mathcal{D}}{t} \right )$ arithmetic operations in the ground
field where $\mathcal{D} = 2(n+s)\ D^s(D-1)^{n-s+1}\
\binom{n}{s}$. In practice, it allows us to solve quantifier
elimination problems which are out of reach of the state-of-the-art
(up to $8$ variables).
\end{abstract}

\setlength{\parindent}{0pt}
\footnotesize{\thanks{Huu Phuoc Le and Mohab Safey El Din are
supported by the ANR grants ANR-18-CE33-0011 \textsc{Sesame}, and
ANR-19-CE40-0018 \textsc{De Rerum Natura}, the joint ANR-FWF
ANR-19-CE48-0015 \textsc{ECARP} project and the European Union's
Horizon 2020 research and innovative training network program under
the Marie Sk\l{}odowska-Curie grant agreement N° 813211 (POEMA).}}

\setlength{\parindent}{20pt}
\section{Introduction}
\label{section:intro}
\paragraph*{Problem statement.}
Let $\ff = (f_1,\ldots,f_s) \subset \QQ[\by][\bx]$ with $\bx = 
(x_1,\ldots,x_n)$ and $\by = (y_1,\ldots,y_t)$.
We aim at solving the following quantifier elimination problem over the 
reals
\[ \exists (x_1, \ldots, x_n)\in \RR^n \quad f_1(\bx,
\by)=\cdots=f_s(\bx, \by)=0.\]
This consists in computing a logically equivalent
\emph{quantifier-free} semi-algebraic formula $\Phi(\by)$, i.e.
$\Phi$ is a finite disjunction of conjonctions of polynomial
constraints in $\QQ[\by]$ which is true if and only if the input
quantified formula is true. The $\bx$ variables are called
\emph{quantified} variables and the $\by$ variables are called
\emph{parameters}.

Let $\pi$ be the projection $(\bx, \by)\mapsto \by$. Note that,
geometrically, $\Phi$ describes the \emph{projection} on the
$\by$-space of the real algebraic set $\cV_{\RR}\subset \RR^{t}\times
\RR^{n}$ defined by simultaneous vanishing of the $f_i$'s. In this
paper, we focus on solving a variant of the classical one block
quantifier elimination, which computes a semi-algebraic formula which
defines a dense subset of the interior of $\pi(\cV_{\RR})$.
\begin{example}
Consider the toy example $x^2+y^2 = 1$.
Its projection on the $y$ coordinate is
described by the quantifier-free formula
$(y \ge -1) \wedge (y \leq
1)$ while for our variant quantifier elimination problem, an admissible 
output is
$(y > -1) \wedge (y < 1)$.
\end{example}
Except for proving theorems, this is sufficient for most of
applications of quantifier elimination in engineering sciences or
computing sciences where either the output formula only needs to
define a sufficiently large subset of the $\pi(\cV_{\RR})$ or is
evaluated with parameters's values which are subject to numerical
noise.
\paragraph*{Prior works.}
The real quantifier elimination is a fundamental problem in
mathematical logic and computational real algebraic geometry. It
naturally arises in many problems in diverse application areas. The
works of Tarski and Seidenberg \citep{Tarski51, Sei54} imply that the
projection of any semi-algebraic set is also semi-algebraic and give
an algorithm, which is however not elementary recursive, to compute
this projection. The Cylindrical Algebraic Decomposition (CAD)
\cite{Collins76} is the first effective algorithm for this problem
whose complexity is doubly exponential in the number of indeterminates
\cite{DavHeintz88}. Since then, there have been extensive researches
on developing this domain. We can name the CAD variants with improved
projections \cite{McCallum88,Hong90,McCallum99,Brown01} or the partial
CAD \cite{ColHo91}. Following the idea of \cite{Gri88} that exploits
the block structure, \cite{Renegar92,BPR96} introduced algorithms of
only doubly exponential complexity in the order of quantifiers (number
of blocks). For one-block quantifier elimination, the arithmetic
complexity and the degree of polynomials in the output of these
algorithms are of order $s^{n+1}D^{O(nt)}$ where $D$ is the bound on
the degree of input polynomials (see \cite[Algo 14.6]{BPR}). However,
obtaining efficient implementations of these algorithms remains
challenging. We also cite here some other works in real quantifier
elimination \cite{Weis88,StWe96,Weis98,BrChris06,Strze06} and
applications to other fields \cite{LiSte93,AnaiWei01,StTiwari11}.

In spite of this tremendous progress, many important
applications stay out of reach of the state-of-the-art of the classic
quantifier elimination. This motivates the researches on its
variants. Generic quantifier elimination, in which the input and
output formulas are equivalent for only almost every parameter, is
studied in \cite{DSWe98,SeiSturm03}. A practically efficient algorithm
is presented in \cite{HoSa09,HoSa12} for the same problem but under
some assumptions on the input. The variant studied in this paper is a
particular instance of the one in \cite{HoSa09, HoSa12}.
\paragraph*{Main results.}
In this paper, we consider the input $\ff=(f_1,\ldots,f_s)$ satisfying
the assumptions below.
\begin{assumption}{A}
  \label{assumption:B2}
  \leavevmode
  \begin{itemize}
    \item The ideal of $\QQ[\bx,\by]$ generated by $\ff$ is radical.
    \item The algebraic set $\cV \subset \CC^{t+n}$ of $\ff$ is
      equi-dimensional of dimension $d+t$. Its
      singular locus has dimension at most $t-1$.
  \end{itemize}
\end{assumption}
\begin{assumption}{B}
  \label{assumption:A}
The Zariski closure $\overline{\pi(\cV)}$ of $\pi(\cV)$ is the whole
parameter space $\CC^t$ and $\pi(\cV_{\RR})$ is not of zero-measure in
$\RR^t$.
\end{assumption}
The first result of the paper is a new probabilistic algorithm for
solving the aforementioned variant of the quantifier elimination on
such an input $\ff$. Our algorithm applies the algorithm of
\cite{SaSc03} to the system $\ff$ considering $\QQ(\by)$ as the based
field. This allows to reduce our problem to zero-dimensional
polynomial systems in $\QQ(\by)[\bx]$. Next, we compute
semi-algebraic formulas that describe approximate projections of these
systems on the $\by$-space through the algorithm of \cite{LS20}. This
algorithm relies on a parametric variant of Hermite matrices for real
root counting \cite{PRS93, Hermite56}. A similar outline is also
presented in \cite{Weis98,Dolz04}, in which the author computes an
expensive comprehensive Gr\"obner bases \citep{Weis92} to analyze all
cases before applying the real root counting algorithm of
\cite{PRS93}. The relaxation of the output allows us to replace this
exhaustive computation by the real root finding algorithm of
\cite{SaSc03}.

Our second goal is to analyze the complexity of this new
algorithm. For generic inputs, we bound the degree of the outputs and
establish an arithmetic complexity which depends on this bound. The
precise notion of genericity is as follows.

Let $\CC[\bx,\by]_{\leq D} = \{p \in
\CC[\bx,\by] \; | \; \deg(p) \leq D\}$. A property $P$ is said to be
generic over $\CC[\bx,\by]_{D}^{s}$ if and only if there exists a
non-empty Zariski open subset $\mathscr{P} \subset
\CC[\bx,\by]_{D}^{s}$ such that the property $P$ holds for every $\ff
\in \mathscr{P}$.

Our complexity result is then stated below. The notation $O\
{\widetilde{~}}(g)$ means $O(g\log^{\kappa}(g))$ for some $\kappa >
0$.
\begin{theorem}
\label{thm:main}
Let $\mathcal{D} = 2(n+s)\ D^s(D-1)^{n-s+1}\ \binom{n}{s}$. There
exists a non-empty Zariski open subset $\mathscr{F}$ of
$\CC[\bx,\by]_{\leq D}^{s}$ and a probabilistic algorithm such that,
for every $\ff \in \mathscr{F}$, this algorithm, in case of success,
computes a semi-algebraic formula $\Phi$ defining a dense subset of
the interior of $\pi(V(\ff)\cap \RR^{t+n})$ within
\[\textstyle{O\ {\widetilde{~}}\left ( (n-s+1)\ 8^{t}\ \mathcal{D}^{3t+2}\
\binom{t+\mathcal{D}}{t} \right )}\]
arithmetic operations in $\QQ$ and $\Phi$ involves only polynomials in
$\QQ[\by]$ of degree at most $\mathcal{D}$.

\end{theorem}
Even though our complexity result has the same order as the one of
\cite[Algo 14.6]{BPR}, we obtain explicitly the degree bounds on the
output formulas and the constant in the $O$ notation in the exponent.

On the practical aspect, our implementation in \textsc{Maple} of this
algorithm outperforms real quantifier elimination functions in
\textsc{Maple} and \textsc{Mathematica}. It allows us to solve 
examples, both generic and non-generic, that are out of reach of these
softwares (up to $8$ indeterminates). These timings are reported in
Section~\ref{section:experiment}.
\paragraph*{Structure of the paper.}
In Section~\ref{section:preliminary}, we start by recalling some basic
notions.  In Section~\ref{section:SaSc03}, we resume the algorithm for
real root finding of \cite{SaSc03}. Also in the same section, we prove
some auxiliary results in order to apply this algorithm
parametrically. Next, we dedicate Section~\ref{section:algorithm} for
the description of our algorithm for solving the targeted problem and
proving its correctness. The complexity of this algorithm is analyzed
in Section~\ref{section:complexity}. Finally, we report on some
experimental results in Section~\ref{section:experiment}.
\section{Preliminaries}
\label{section:preliminary}
\paragraph{Algebraic sets and critical points.}
Let $\FF$ be a subfield of $\CC$ and $F \subset
\FF[x_1,\ldots,x_n]$. The algebraic subset of $\CC^n$ at which the
elements of $F$ vanish is denoted by $\VV(F)$. For an algebraic set
$\cV\subset \CC^n$, we denote by $I(\cV)\subset \CC[x_1,\ldots,x_n]$
the radical ideal associated to $\cV$. The singular locus of $\cV$ is
denoted by ${\rm sing}(\cV)$. Given any subset $\mathcal{S}$ of
$\CC^n$, we denote by $\overline{\mathcal{S}}$ the Zariski closure of
$\mathcal{S}$, i.e., the smallest algebraic set containing
$\mathcal{S}$. An algebraic set $\cV$ is equi-dimensional if its
irreducible components share the same dimension.

A map $\varphi$ between two algebraic sets $\cV \subset \CC^n$ and
$\cW\subset \CC^i$ is a polynomial map if there exist
$\varphi_1,\ldots,\varphi_i \in \CC[x_1,\ldots,x_n]$ such that
$\varphi(\bma)=(\varphi_1(\bma),\ldots,\varphi_i(\bma))$ for $\bma\in
\cV$. Let $\cV\subset \CC^n$ be an equi-dimensional algebraic set. We
denote by $\crit(\varphi, \cV)$ the set of critical points of the
restriction of $\varphi$ to the non-singular locus of $\cV$. If $c$ is
the codimension of $\cV$ and $(f_1,\ldots,f_s)$ generates the ideal
$I(\cV)$, the subset of $\cV$ at which the Jacobian matrix $\jac(f_1,
\ldots, f_s, \varphi_1, \ldots, \varphi_i)$ of $(f_1,
\ldots, f_s, \varphi_1,\ldots, \varphi_i)$ has rank less than or equal
to $c$ is the union of $\crit(\varphi, \cV)$ and $\sing(\cV)$ (see,
e.g., \cite[Subsection 3.1]{SaSc17}).
\paragraph{Gr\"obner bases and zero-dimensional ideals.} Let
$\mathbb{F}$ be a field and $\overline{\mathbb{F}}$ be its algebraic
closure. We fix an admissible monomial order $\succ$ (see
\cite[Sec. 2.2]{CLO}) over $\mathbb{F}[\bx]$ where
$\bm{x}=(x_1,\ldots,x_n)$. For $p\in \mathbb{F}[\bm{x}]$, the leading
monomial of $p$ with respect to $\succ$ is denoted by
$\lm_{\succ}(p)$.

A Gr\"obner basis $G$ of an ideal $I\subset\mathbb{F}[\bm{x}]$ w.r.t.
the order $\succ$ is a finite generating set of $I$ such that the set
of leading monomials $\{ \lm_{\succ}(g) \; | \; g\in G\}$ generates
$\langle \lm_{\succ}(p) \; |\; p\in I\rangle$. For $p \in
\mathbb{F}[\bm{x}]$, the remainder of the division of $p$ by $G$ using
the order $\succ$ is uniquely defined and is called the {\em normal
form} of $p$ w.r.t. $G$. A polynomial $p$ is reduced by $G$ if $p$
equals to its normal form w.r.t. $G$.

An ideal $I$ is said to be zero-dimensional if the algebraic set $V(I)
\subset \overline{\FF}^n$ is finite and non-empty. When this holds, by
\cite[Sec.  5.3, Theorem 6]{CLO}, the quotient ring
$\mathbb{F}[\bm{x}]/I$ is a $\mathbb{F}$-vector space of finite
dimension. The dimension of this vector space is also called the
algebraic degree of $I$; it coincides with the number of points of
$V(I)$ counted with multiplicities \cite[Sec. 4.5]{BPR}. For any
Gr\"obner basis of $I$, the set of monomials in $\bm{x}$ which are
irreducible by $G$ forms a monomial basis, denoted by $B$, of this
vector space. For $p\in \mathbb{F}[\bx]$, the normal form of $p$ by
$G$ can be interpreted as its image in $\FF[\bx]/I$ and is a
$\mathbb{F}$-linear combination of elements of $B$.
\paragraph{Properness \& Noether normalization.} A map $\varphi: V
\mapsto \CC^i$ is proper at $\beta \in \CC^i$ if there exists a
neighborhood $\mathcal{O}$ of $\beta$ such that
$\varphi^{-1}(\overline{\mathcal{O}})$ is compact, where
$\overline{\mathcal{O}}$ denotes the closure of $\mathcal{O}$ in the
Euclidean topology. If $\varphi$ is proper everywhere on its image, we
say that the map $\varphi$ is proper.  The properness is
strongly related to the following notion of Noether normalization.

Let $\FF$ be a field and $I$ be an ideal of $\FF[x_1,\ldots,x_n]$. The
variables $(x_{i+1},\ldots,x_n)$ are in Noether position w.r.t. $I$ if
their canonical images in the quotient algebra
$\FF[x_1,\ldots,x_{n}]/I$ are algebraic integers over $\FF[x_{1},
\ldots, x_{i}]$ and $\FF[x_{1},\ldots,x_{i}] \cap I = \langle 0
\rangle$. Once $\FF = \CC$ and the variables $(x_{i+1},\ldots,x_n)$ is
in Noether position w.r.t. $I$, the projection of $V(I)$ on
$(x_1,\ldots,x_i)$ is proper.
\paragraph{Change of variables.}
Given a field $\FF$, we denote by $\GL(n, \FF)$ the set of invertible
matrices of size $n\times n$ with entries in $\FF$. Let $p \in
\FF[\bx]$ be a polynomial.  For any $A\in \GL(n, \FF)$, we denote by
$p^{A}$ the polynomial $p(A\cdot \bx) \in \FF[\bx]$. For any algebraic
set $V \subset \overline{\FF}^n$, $V^A$ denotes the algebraic set
$\{A^{-1} \cdot \bx \; | \; \bx \in V \}$.

For two blocks of indeterminates $\bx$ and $\by$, we consider
frequently the matrices that act only on the variables $\bx$ and leave
$\by$ invariant. Those matrices form a subset denoted by ${\rm
GL}(n,t,\FF)$ of ${\rm GL}(n+t,\FF)$.


\section{Algorithm for real root finding}
\label{section:SaSc03}
\subsection{The $S^2$ algorithm}
We recall the algorithm in \cite{SaSc03}, which we refer to as the $S^2$
algorithm, that computes at least one point per connected component of
a smooth real algebraic set.

Let $\ff = (f_1,\ldots,f_s)$ be a polynomial sequence in
$\RR[x_1,\ldots,x_n]$. For $1\leq i \leq d$, let $\phi_i$ be the
projection $(x_1, \ldots, x_n)\to (x_1, \ldots, x_i)$. When $\ff$
defines a smooth equi-dimensional algebraic set $\cV \subset \CC^n$
and generates a radical ideal, one can build a polynomial system using
appropriate minors of $\jac(\ff)$ to define $\crit(\phi_i, \cV)$. Note
that the critical loci are nested
\[ \crit(\phi_1, \cV)\subset \crit(\phi_2, \cV)\subset \cdots
\crit(\phi_d, \cV) \subset \crit(\phi_{d+1}, \cV) = \cV.\]
Note also that in \emph{generic} coordinates $\crit(\phi_i, \cV)$ has 
expected dimension $i-1$.  The algorithm in \cite{SaSc03} then exploits 
stronger properties of these critical loci under some genericity
assumption on the coordinate system (which are satisfied through a
generic linear change of coordinates).
\begin{proposition}{\cite[Theorem 2]{SaSc03}}
\label{prop:genvars}
Assume that $\ff$ defines a smooth equi-dimensional algebraic set and
generates a radical ideal.
    
Then, there exists a non-empty Zariski open set $\zarA_{\ff} \in
\GL(n, \CC)$ such that for $A \in \zarA_{\ff}$ the following holds:
  \begin{itemize}
    \item the restriction of $\phi_{i-1}$ to $\crit(\phi_i,
\cV^A)$ is proper;
    \item the set $\crit(\phi_i, \cV^A)$ is either empty or of
dimension $i-1$ for $1 \leq i \leq d+1$.
    \end{itemize}
  \end{proposition}
The first item in Proposition~\ref{prop:genvars} implies
the second one. The index in the notation $\zarA_{\ff}$ indicates that
the non-empty Zariski open set depends on $\ff$.  Algorithm $S^2$
considers fibers of the above critical loci with the convention
$\pi_0: \bm{x}\to \bullet$.  Proposition~\ref{prop:genvars} is the
cornerstone of the $S^2$ algorithm which can be derived from the
following one.

\begin{proposition}{\cite[Theorem 2]{SaSc03}}\label{prop:sampling}
Assume that $\ff$ defines a smooth equi-dimensional algebraic set and
generates a radical ideal.

For $A\in \zarA_{\ff}\cap \GL(n, \QQ)$ as defined in
Proposition~\ref{prop:genvars} and $\alpha =
(\alpha_1,\ldots,\alpha_d) \in \RR^d$, the union of the sets
\[\crit(\phi_i, \cV^{A})\cap
\phi_{i-1}^{-1}((\alpha_1,\ldots,\alpha_{i-1})), \quad 1 \le i \le d+1
\]
is finite and meets all connected components of $\cV\cap\RR^n$.
\end{proposition}
\begin{example}
Let $\cV$ be the smooth surface defined by
$x_1^2-x_2^2-x_3^2 = 1$. The Jacobian matrix $\jac(\ff)$
writes simply $(2x_1,-2x_2,-2x_3)$. It turns out that the identity
matrix lies in the set $\zarA$ defined in
Proposition~\ref{prop:genvars}. Taking $\alpha = (0,0)$, we obtain
3 zero-dimensional systems:
\begin{itemize}
\item $\crit(\phi_1,\cV)$: $\{-2x_2, -2x_3, x_1^2-x_2^2-x_3^2 - 1\}$,
\item $\crit(\phi_2,\cV)\cap \phi_1^{-1}(\bm{0})$: $\{-2x_3,
  x_1^2-x_2^2-x_3^2-1,x_1\}$,
\item $\cV \cap \phi_2^{-1}(\bm{0})$: $\{x_1^2-x_2^2-x_3^2-1,x_1,x_2\}$.
\end{itemize}
The first system admits two real solutions $(1,0,0)$ and $(-1,0,0)$. The
other systems do not have any real solution. The two points $(1,0,0)$
and $(-1,0,0)$ intersect the two connected components of $\cV$.

Of course, on general examples, one would need to perform a randomly
chosen linear change of variables but this example illustrates already
how $S^2$ works.
\end{example}

\subsection{Parametric variant of $S^2$}
We present now a parametric variant of $S^2$.  We let $\ff =
(f_1,\ldots,f_s) \subset \QQ[\by][\bx]$ where $\by = (y_1,\ldots,y_t)$
are considered as parameters and $\bx = (x_1,\ldots,x_n)$ are
variables. The algebraic set defined by $\ff$ is denoted by $\cV
\subset \CC^t\times\CC^n$. Let $\pi$ denote the projection $(\bx,\by)
\mapsto \by$ and $\pi_i$ denote the projection $(\by,\bx) \mapsto
(\by,x_1,\ldots,x_i)$.

Considering $\QQ(\by)$ as the ground field, the parametric variant of
$S^2$ computes on the input $\ff$ a list of finite subsets of
$\QQ[\by][\bx]$, each of which generates a zero-dimensional ideal of
$\QQ(\by)[\bx]$. These subsets are basically $\ff^{A} \cup
\Delta_i^{A} \cup \{x_1-\alpha_1,\ldots,x_{i-1}-\alpha_{i-1}\}$, where
$(A,\alpha)$ is randomly chosen in ${\rm GL}(n,t,\QQ)\times \QQ^n$ and
$\Delta_i^{A}$ is the set of all $(n-d)$-minors of the Jacobian matrix
of $\ff^{A}$ w.r.t. $x_{i}, \ldots, x_{n}$.

The rest of this subsection is devoted to the auxiliary results that
allow us to use the $S^2$ algorithm parametrically as above.
\begin{lemma}
\label{lemma:B-param}
When Assumptions~\eqref{assumption:B2} and~\eqref{assumption:A} hold,
there exists a non-empty Zariski open subset $\mathscr{B}$ of $\CC^t$
such that for every $\eta \in \mathscr{B}$, the specialization
$\ff(\eta,\cdot)$ of $\ff$ at $\eta$ generates a radical
equi-dimensional ideal whose algebraic set is either empty or has
dimension $d$.
\end{lemma}
\begin{proof}
Under Assumption~\eqref{assumption:A}, by the fiber dimension theorem
\cite[Theorem 1.25]{Shafa}, there exists a non-empty Zariski open
subset $\mathscr{B}'$ of $\CC^t$ such that $\pi^{-1}(\eta)\cap \cV$ is
an algebraic set of dimension $d$.

Let $\mathcal{W}$ denote the set of points of $\cV$ at which the
Jacobian matrix $\jac_{\bx}(\ff)$ of $\ff$ w.r.t. $\bx$ has rank at most
$n-d-1$. We note that $\mathcal{W} = \crit(\pi,\cV) \cup {\rm
sing}(\cV)$. The algebraic version of Sard's theorem \cite[Proposition
B2]{SaSc17} implies that $\pi(\crit(\pi,\cV))$ is contained in a
proper Zariski closed subset of $\CC^t$. On the other hand, as
Assumptions~\eqref{assumption:B2} hold, the dimension of $\pi\left(
\sing(\cV) \right)$ is less than $t$. Thus, it is also contained in a
proper Zariski closed subset of $\CC^t$.

Hence, the Zariski closure of $\pi(\mathcal{W})$ is a proper Zariski
closed subset of $\CC^t$. Let $\mathscr{B}$ be the intersection of the
complement in $\CC^t$ of this Zariski closure with $\mathscr{B}'$. For
$\eta \in \mathscr{B}$, the set
\[\{\bx \in \CC^n \; | \; \ff(\eta, \bx) = 0,
  \rank\jac_{\bx}(\ff)(\eta) < n -d \}\]
is empty. Since the dimension
of $\pi^{-1}(\eta)\cap \cV$ is $d$ and the Jacobian matrix
$\jac_{\bx}(\ff)(\eta, \cdot)$ of $\ff(\eta,\cdot)$ w.r.t. the
variables $\bx$ is of rank $n-d$ for every $(\eta,\bx) \in \cV\cap
\pi^{-1}(\eta)$, the ideal $\ff(\eta,\cdot)$ is radical and defines a
smooth and equi-dimensional set of dimension $d$ by Jacobian criterion
\cite[Theorem 16.19]{Eisenbud}.
\end{proof}
Lemma~\ref{lemma:B-param} shows that when specializing $\by = (y_1,
\ldots, y_t)$ to a generic point $\eta\in \mathscr{B}\cap \RR^t$ in
$\ff$, one obtains $\ff(\eta,\cdot)$ satisfying the assumptions of
Proposition~\ref{prop:genvars}. One could then apply Algorithm $S^2$
to $\ff(\eta,\cdot)$ to grab sample points in the real algebraic set
it defines. However, proceeding this way would lead us to use a change
of variables encoded by a matrix $A$ depending on $\eta$. The result
below shows that choosing one generic change of variables
will be valid for most of parameters' values.
\begin{proposition}
  \label{proposition:Noether-parametric}
Assume that Assumptions~\eqref{assumption:B2} and~\eqref{assumption:A}
hold. There exists a dense Zariski open subset $\mathscr{O}$ of
${\rm GL}(n,t,\CC)$ such that for every $A \in \mathscr{O}\cap {\rm
GL}(n,t,\QQ)$ the following holds.

There exists a dense Zariski open subset $\mathscr{Y}_{A}$ of $\CC^t$
such that $\mathscr{Y}_{A}$ is a subset of the Zariski open set
$\mathscr{B}$ in Lemma~\ref{lemma:B-param} and $A$ lies in the Zariski
open set $\zarA_{\ff(\eta, .)}$ defined in
Proposition~\ref{prop:genvars} for every $\eta \in \mathscr{Y}_{A}$.
\end{proposition}

\begin{proof}
Let $\overline{\CC(\by)}$ denote the algebraic closure of
$\CC(\by)$. We consider $\overline{\CC(\by)}$ as the coefficient
field. The proof of \cite[Theorem 1]{SaSc03} is purely algebraic and
then is valid over the based field $\overline{\CC(\by)}$. Hence, there
exists a non-empty Zariski open subset $\tilde{\mathscr{O}}$ of ${\rm
GL}(n,t,\overline{\CC(\by)})$ such that for $A \in
\tilde{\mathscr{O}}\cap {\rm GL}(n,t,\QQ)$, the variables
$(x_1,\ldots,x_{i-1})$ is in Noether position w.r.t. the ideal in
$\QQ(\by)[\bx]$ generated by $\ff^{A} + \Delta_{i}^{A}$ for $1\leq i
\leq d+1$ where $\Delta_i^A$ is the set of maximal minors of the
truncated Jacobian matrix of $\jac(\ff^A)$ with all the partial
derivatives w.r.t.  $\by$ and $x_j$ for $1\leq j \leq i$ being removed
(hence these minors are the ones defining $\crit(\pi_i,\cV) \cup
\sing(\cV)$).

This is equivalent to the following. For $1\leq i \leq d+1$, $i \leq j
\leq n$, there exist the polynomials $p_{i,j} \in
\QQ(\by)[x_1,\ldots,x_{i-1},x_j]$ such that each $p_{i,j}$ lies in the
ideal of $\QQ(\by)[\bx]$ generated by $\ff^{A} \cup \Delta_{i}^{A}$
and it is monic when considering $x_j$ as the only variable (with the
coefficients in $\QQ(\by)[x_{1},\ldots,x_{i-1}]$).

The denominators of $p_{i,j}$ are then polynomials in $\QQ[\by]$. We
choose $\mathscr{Y}_A$ to be the intersection of the non-empty Zariski
open set $\mathscr{B}$ defined in Lemma~\ref{lemma:B-param} and the
non-empty Zariski open set defined by the non-vanishing of all the
denominators appeared in the $p_{i,j}$'s. Thus, for $\eta \not \in
\mathscr{Y}_A$, $p_{i,j}(\eta,\cdot) \in \QQ[x_1,\ldots,x_{i-1},x_j]$
is monic in $x_j$. Consequently, $(x_i,\ldots,x_{n})$ is in Noether
position w.r.t. the ideal of $\CC[\bx]$ generated by $\ff^A(\eta,\cdot)
\cup \Delta_i^A(\eta,\cdot)$. Finally, taking $\mathscr{O} =
\tilde{\mathscr{O}} \cap {\rm GL}(n,t,\CC)$, the conclusion follows.
\end{proof}

\section{One-block QE algorithm}
\label{section:algorithm}
\subsection{Description}
In this subsection, we describe our algorithm for solving our variant
of the quantifier elimination problem. The input is a polynomial
sequence $\ff = (f_1,\ldots,f_s) \subset \QQ[\bx,\by]$ satisfying
Assumptions~\eqref{assumption:B2} and \eqref{assumption:A}. Further,
we denote by $Z(\Psi)$ the zero set of any semi-algebraic formula
$\Psi$, i.e., $Z(\Psi) = \{\by \in \RR^t \; | \; \Psi(\by) \text{ is
true}\}$.

By Assumptions~\eqref{assumption:B2} and~\eqref{assumption:A}, the
fiber dimension theorem \cite[Theorem 1.25]{Shafa} implies that there
exists a non-empty Zariski open subset of $\CC^t$ such that
$\pi^{-1}(\eta)$ has dimension $d$. The idea is to apply the parametric 
variant of $S^2$ with $\QQ(\by)$ as a ground field. 

More precisely, we start by picking randomly $(A,\alpha)$ in
${\GL}(n,t,\QQ) \times \QQ^n$ and apply the change of variables $\bx
\mapsto A\cdot \bx$ to the input $\ff$ to obtain a new sequence
$\ff^{A}$. As $A$ acts only on $\bx$, $\pi(V(\ff^{A})\cap \RR^{n+t}) =
\pi(\cV_{\RR})$. Hence, a quantifier-free formula that solves our
problem for $\ff^{A}$ is also a solution of the same problem for
$\ff$.

Let $\jac_{\bx}(\ff^{A})$ be the Jacobian matrix of $\ff^{A}$ w.r.t.
the variables $\bx = (x_1,\ldots,x_n)$. We denote by $J_1,\ldots,J_n$
the columns of $\jac_{\bx}(\ff^{A})$ respectively. For $1\leq i \leq
d$, let $W_i^{A,\alpha}$ be the union of $\ff^{A}$, all
the $(n-d)$-minors of the matrix consisting of the columns
$J_{i+1},\ldots,J_n$ and
$\{x_1-\alpha_1,\ldots,x_{i-1}-\alpha_{i-1}\}$. In particular,
$W^{A,\alpha}_{d+1}$ denotes $\ff^{A}\cup
\{x_1-\alpha_1,\ldots,x_{d}-\alpha_d\}$.

We prove later in Lemma~\ref{lemma:dimension-parametric} that, for
generic $(A,\alpha)$, the ideals of $\QQ(\by)[\bx]$ generated by
$W_i^{A,\alpha}$ are radical and zero-dimensional.

We now solve the quantifier elimination problem for each of the
polynomial sets $W_i^{A,\alpha}$. For this step, we refer to a subroutine
called {\sf RealRootClassification} that takes as input a polynomial
sequence $F \subset \QQ[\by][\bx]$ such that the ideal of
$\QQ(\by)[\bx]$ generated by $F$ is radical and zero-dimensional and
computes a quantifier-free formula $\Phi_F$ in $\by$
such that $Z(\Phi_{F})$ is dense in the interior of $\pi(V(F)\cap
\RR^{n+t})$. For this task, we refer to the algorithm of
\cite{LS20}. We will explain the essential details of this subroutine
later in Subsection~\ref{ssection:rrc}.

Calling the subroutine {\sf RealRootClassification} on the inputs
$W_i^{A,\alpha}$ gives us the lists of semi-algebraic formulas
$\Phi_i$. Finally, we return $\Phi = \vee_{i=1}^{d+1} \Phi_i$ as the
output of our algorithm.

The pseudo-code below summarizes our algorithm, we introduce two
additional subroutines:
\begin{itemize}
  \item {\sf GenericDimension} which takes the sequence $\ff$
and computes the dimension of the ideal generated by $\ff$ in
$\QQ(\by)[\bx]$.
  \item $(n-d)\ \mathsf{Minors}$ which takes as input a matrix $M$
whose coefficients are in $\QQ[\bx,\by]$ and computes all of its
$(n-d)$-minors.
\end{itemize}
\begin{algorithm}
\small
\DontPrintSemicolon
\KwData{$\ff \in \QQ[\by][\bx]$ satisfying
Assumptions~\eqref{assumption:B2} and~\eqref{assumption:A}.}
\KwResult{A formula $\Phi$ s.t $Z(\Phi)$ is dense in the interior of
$\pi(\cV_{\RR})$.}
Choose randomly $(A,\alpha) \in \mathrm{GL}(n,\QQ) \times \QQ^n$ \\
$\ff^{A} \gets \ff(A \cdot \bx)$ \\
$[J_1,\ldots,J_n] \gets \jac_{\bx}(\ff^{A})$ \\
$d \gets {\sf GenericDimension}(\ff^{A})$\\
\For{$1 \leq i \leq d+1$}{
\mbox{$W_i^{A,\alpha} \gets (n-d)\
\mathsf{Minors}([J_{i+1},\ldots,J_{n}]) \cup \{\ff^{A},x_1 - \alpha_1,
\ldots, x_{i-1} - \alpha_{i-1}\}$ }\\
$\Phi_i \gets {\sf RealRootClassification}(W_i^{A,\alpha})$ \\
}
\Return $\Phi \gets \vee_{i=1}^{d+1} \Phi_i$
\caption{{\sf One-block quantifier elimination}}
\label{algo:QE}
\end{algorithm}
\vspace{-0.2cm}
\subsection{Real root classification}
\label{ssection:rrc}
Now we explain the general ideas of the algorithm presented in
\cite{LS20} that is used in the \textsf{RealRootClassification}
subroutine.

Let $F \subset \QQ[\by][\bx]$ be a polynomial sequence such that the
ideal $\langle F \rangle$ generated by $F$ in $\QQ(\by)[\bx]$ is
radical and zero-dimensional.

For such an input $F$, {\sf RealRootClassification} computes a
semi-algebraic formula $\Phi_{F}$ and a polynomial $w_{\infty} \in
\QQ[\by]$ that satisfies:
\begin{itemize}
\label{item:rrc}
  \item $Z(\Phi_{F}) \subset \pi(V(F)\cap \RR^{n+t})$,
  \item $Z(\Phi_{F}) \setminus V(w_{\infty}) = \pi(V(F)\cap \RR^{n+t})
\setminus V(w_{\infty})$.
\end{itemize}
The algorithm in \cite{LS20} is based on constructing a symmetric
matrix $H_F$ with entries in $\QQ(\by)$ associated to $F$. This matrix
is basically a parametric version of the classical Hermite matrix for
the ideal $\langle F\rangle$ (see, e.g., \cite[Chap. 4]{BPR}), which
provides the number of distinct real/complex solutions of the system
$F(\eta,\cdot)$ through the signature/rank of the specialization of
$H_{F}$ at $\eta$ \citep[Corollary 17]{LS20}.

Let $G_F$ be the reduced Gr\"obner basis of the ideal in
$\QQ[\bx,\by]$ generated by $F$ w.r.t. the $\DRL(\bx) \succ \DRL(\by)$
order. We consider the leading coefficients of the elements of $G_F$
in variables $\bx$ w.r.t. the $\DRL(\bx)$ order, which are polynomials
in $\QQ[\by]$. Then, $w_{F}$ is taken as the square-free part of
the product of these leading coefficients. The polynomial $w_{F}$
defines a proper algebraic subset of $\by$-space over which the matrix
$H_F$ does not have good specialization property (\cite[Proposition
16]{LS20}).

Next, we choose randomly a matrix $Q \in {\GL}({\delta},\QQ)$. As the
entries of $H_F$ lie in $\QQ(\by)$, so do the leading principal minors
$M_1,\ldots,M_{\delta}$ of $Q^T\cdot H_F \cdot Q$. Let
$m_1,\ldots,m_{\delta}$ be the numerators of those minors, which are
in $\QQ[\by]$. A sufficiently generic matrix $Q$ ensures that none of
the $m_i$'s is identically zero, hence allowing us to determine the
signature of $H_F$ according to the signs of the $m_i$'s.  We then
compute a finite set of points $L$ of $\QQ^t$ that intersects every
connected component of the semi-algebraic set defined by
$\wedge_{i=1}^{\delta} (m_i \ne 0) \wedge (w_{F} \ne 0)$. Over
those connected components, the polynomials $m_i$ are
sign-invariant. Since the signature of $H_F(\eta)$ can be deduced from
the signs of the $m_i(\eta)$, the number of real solutions of
$F(\eta,\cdot)$ is also invariant when $\eta$ varies in each connected
component.

Let $L_0 = \{\eta \in L \; | \; F(\eta,\cdot) \text{ admits at
least one real solution}\}$ and
\[\Phi_{F} = \left ( \vee_{\eta \in L_0}
    \left(\wedge_{i=1}^{\delta} \sign M_i =
      \sign M_i(\eta) \right) \right) \wedge (w_{F} \ne
  0).\]
Then, $w_{\infty}$ is taken as the product of the $m_i$'s and
$w_{F}$. We return $\Phi_{F}, w_{\infty}$ as the output of {\sf
RealRootClassification} for $F$. The correctness of this algorithm is
given in \cite[Proposition 28]{LS20}.

In the pseudo-code below, we introduce the subroutines
\begin{itemize}
  \item {\sf HermiteMatrix} which takes as input a polynomial sequence
$F \subset \QQ[\by][\bx]$ such that the ideal $\langle F \rangle
\subset \QQ(\by)[\bx]$ is zero-dimensional and computes the parametric
Hermite matrix associated to $F$ w.r.t. the $\DRL(\bx)$ order.

The description of this subroutine is given in \cite[Algo. 2]{LS20}.
  \item {\sf PrincipalMinors} computes the leading principal
minors of the matrix $Q^T \cdot H_{F} \cdot Q$.
  \item {\sf SamplePoints} which takes as input a polynomial sequence
    $m_1,\ldots,m_{\delta}, w_{F} \in \QQ[\by]$ and computes a
    finite set of points that intersects every connected component of
    the semi-algebraic set defined by $\wedge_{i=1}^{\delta}
      m_i \ne 0 \wedge w_{F}\ne 0$.
    
    We describe such a subroutine in \cite[Sec. 3]{LS20}.
  \item {\sf Signature} which evaluates the signature of a symmetric
    matrix of entries in $\QQ$.
\end{itemize}
\begin{algorithm}
\small
\DontPrintSemicolon
\KwData{A polynomial sequence $F \subset \QQ[\by][\bx]$ such that the
ideal of $\QQ(\by)[\bx]$ generated by $F$ is radical and
zero-dimensional.}
\KwResult{A formula $\Phi_{F}$ and a polynomial $w_{\infty} \in
\QQ[\by]$.}
$H_F,w_{F} \gets \textsf{HermiteMatrix}(F)$\\
Choose randomly $Q \in {\rm GL}(\delta,\QQ)$
\tcp{$\delta$ is the size of $H_F$}
$(M_1,\ldots,M_{\delta}) \gets {\sf
PrincipalMinors}(Q^T\cdot H_F \cdot
Q)$ \\
$(m_1,\ldots,m_{\delta}) \gets {\sf
Numerators}(M_1,\ldots,M_{\delta})$ \\
$L \gets {\sf SamplePoints}\left( \left(
  \wedge_{i=1}^{\delta} m_i \ne 0 \right) \wedge w_{F} \ne 0 \right)$\\
\For{$\bma \in L$}{
  \If{${\sf Signature}\left( H_{F}(\bma) \right) \ne 0$}{
$\Phi_F \gets \Phi_F \vee \left( \wedge_{i=1}^{\delta}  \sign
      M_i = \sign M_i(\eta) \right) $}
}
$\Phi_F \gets \Phi_F \wedge (w_{F} \ne 0)$ \\
$w_{\infty} \gets w_{F} \cdot \prod_{i=1}^{\delta} m_i$ \\
\Return $\Phi_F, w_{\infty}$
\caption{{\sf RealRootClassification}}
\label{algo:RRC-Hermite}
\end{algorithm}
We end this subsection by an example to illustrate our algorithm.
\begin{example}
We consider the polynomial $f=x_1^2+y_1x_2^2+y_2x_2+y_3$. Let $\Delta
=y_2^2 -4y_1y_3$. The projection of $V(f) \cap \RR^{5}$ on
$(y_1,y_2,y_3)$ is
\[(\Delta \ge 0 \wedge\ y_1 > 0) \vee (y_1 <0) \vee \left (y_1 = 0
\wedge\left ( (y_2\ne 0 ) \vee (y_2 = 0 \wedge y_3 \leq 0) \right)
\right).\]
Applying the parametric variant of $S^2$ for $A = I_3$ and
$\alpha = (0,0)$, we obtain 2 systems $W_{1} = \{2y_1x_2+y_2, f \}$
and $W_{2} = \{f, x_1\}$. Next, we call {\sf RealRootClassification}
on these systems, choosing $Q=I_2$ to simplify the calculation. We
obtain then $w_{1,\infty} = w_{2,\infty} = y_1$ and the Hermite
matrices:
\[\textstyle{H_1 = \begin{pmatrix}
    2 & 0 \\
    0 & -2y_3 + y_2^2/(2y_1)
  \end{pmatrix}, \quad  H_2=
  \begin{pmatrix}
    2 & -y_2/y_1 \\
    -y_2/y_1 & (-2 y_1 y_3 + y_2^2)/y_1^2
  \end{pmatrix}}.\]
The sequences of leading principal minors are respectively
$[2,\Delta/y_1]$ and $[2,\Delta/y_1^2]$. We compute then $4$
points representing $4$ connected components of the semi-algebraic set
defined by $ y_1 \ne 0 \wedge\ \Delta \ne 0$:
\[(1, 1/8, 0), (-1, 1/8, 0), (1, 1/8, 1/128), (-1, 1/8, -1/128).\]
The matrix $H_2$ has non-zero signature over the first and second
points, which both lead to the sign condition $\Delta > 0 \wedge y_1^2
> 0$. Thus, we have
\[\Phi_2 = (\Delta > 0 \wedge y_1^2 > 0) \wedge (y_1\ne 0).\]
For $H_{1}$, non-zero signatures are satisfied at the first and fourth
points. Evaluating the sign of $\Delta$ and $y_1$ at those points
gives
\[\Phi_1 = \left( (\Delta > 0 \wedge\ y_1 > 0) \vee (\Delta < 0
\wedge\ y_1 < 0) \right) \wedge (y_1 \ne 0).\]
The final output is therefore $\Phi = \Phi_1 \vee \Phi_2$, which is
equivalent to
\begin{align*}
\Phi & = (\Delta > 0 \wedge y_1 > 0) \vee (\Delta < 0 \wedge y_1 < 0)
\vee (\Delta > 0 \wedge y_1\ne 0) \\
& = (\Delta > 0 \wedge y_1 > 0) \vee (\Delta \ne 0
\wedge y_1 < 0).\end{align*}
It is straight-forward to see that $Z(\Phi)$ is a
dense subset of $\pi\left (V(f) \cap \RR^{5} \right )$.
\end{example}
\subsection{Correctness of Algorithm~\ref{algo:QE}}
We start by proving that the polynomial sequences $W_i^{A,\alpha}$
satisfy the assumptions required by {\sf RealRootClassification}.

\begin{lemma}
\label{lemma:dimension-parametric}
Assume that Assumptions~\eqref{assumption:B2} and \eqref{assumption:A}
hold. Let $\mathscr{O}$ be the Zariski open subset of ${\rm
GL}(n,t,\CC)$ defined in
Proposition~\ref{proposition:Noether-parametric} and $A\in \mathscr{O}
\cap {\GL}(n,t,\QQ)$. There exists a non-empty Zariski open subset
$\mathscr{X}$ of $\CC^d$ such that for $\alpha \in \mathscr{X} \cap
\QQ^d$, the ideal of $\QQ(\by)[\bx]$ generated by $W_i^{A,\alpha}$ is
radical and either empty or zero-dimensional.
\end{lemma}
\begin{proof}
By Proposition~\ref{proposition:Noether-parametric}, the algebraic set
defined by $W_i^{A,\alpha}(\eta,\cdot)$ is finite when $\eta$ varies
over a non-empty Zariski open subset $\mathscr{Y}_A$ of $\CC^t$. Thus,
the ideal of $\QQ(\by)[\bx]$ generated by $W_i^{A,\alpha}$ is
zero-dimensional. Now we prove that the ideal generated by
$W_i^{A,\alpha}$ is radical.

Let $M_1^A, \ldots, M_{\ell}^A$ be the $(n-d)$ minors of the Jacobian
matrix $J$ associated to $\ff^A$ when considering only the partial
derivatives w.r.t.  $x_{i+1}, \ldots, x_n$. Recall that
$W_i^{A,\alpha}$ is the union of $\ff^A$ with the $M^A_1, \ldots,
M^A_\ell$ with $x_1-\alpha_1, \ldots, x_{i-1}-\alpha_{i-1}$. Further,
we denote by ${W'}_i^A\subset \QQ(\by)[\bx]$ the ideal generated by
$\ff^A, M^A_1, \ldots, M^A_\ell$.

The idea is to follow \cite[Definitions 3.2 and 3.3]{SaSc17} where 
\emph{charts} and \emph{atlases} are defined for algebraic sets defined by 
the vanishing of $\ff^A$ and  $M^A_1, \ldots, M^A_\ell$.  

Let $m$ be a $(n-d-1)$ minor of $J$.  Without loss of generality we
assume that it is the upper left such minor and let $M_1^A, \ldots,
M_{d-(i-1)}^A$ be the $(n-d)$ minors of $J$ obtained by completing $m$
with the $n-d$-th line of $J$ and the missing column.  We denote by
$\QQ(\by)[\bx]_m$ the localized ring where divisions by powers of $m$
are allowed.

By \cite[Lemma B.12]{SaSc17} there exists a non-empty Zariski open set 
$\mathscr{O}'_{m, n-d}$ such that for  $A\in \GL(n,t,\CC)$, the 
localization of the ideal generated by $f_1^A, \ldots, f_{n-d}^A, M_1^A, 
\ldots, M_{d-(i-1)}^A
$ in the ring $\QQ(\by)[\bx]_{m}$ is radical and coincides with the 
localization of ${W'}_i^A$ in $\QQ(\by)[\bx]_{m}$.
By \cite[Prop. 3.4]{SaSc17}, there exists a non-empty Zariski open set
$\mathscr{O}''\subset \GL(n, t, \CC)$ such that for $A \in
\mathscr{O}''$, any irreducible component of the algebraic set defined
by ${W'}_i^A$ contains a point at which a $(n-d-1)$ minor of $J$ does
not vanish.  This implies that any primary component ${W'}_i^A$ whose
associated algebraic set contains such a point is radical and then
prime.

Now define $\Omega$ as the intersection of $\mathscr{O}$ (defined in
Proposition~\ref{proposition:Noether-parametric}), all non-empty
Zariski open sets $\mathscr{O}'_{m, k}$ and $\mathscr{O}''$. Hence,
we then deduce that ${W'}_i^A$ generates a radical ideal. It remains to prove that there exists a non-empty Zariski open set
$\mathscr{X}_i \subset \CC^{i-1}$ such that for $\alpha = (\alpha_1,
\ldots, \alpha_{i-1}) \in \mathscr{X}_i$, $\langle {W'}_i^A\rangle +
\langle x_1 - \alpha_1, \ldots, x_{i-1} - \alpha_{i-1}\rangle$ is
radical in $\QQ(\by)[\bx]$. Choosing $\alpha$ outside the set of
critical values of $\pi_i$ restricted to the algebraic set defined by
${W'}_i^A$ in $\overline{\QQ(\by)}^n$ is enough. By Sard's theorem,
this set of critical values is contained in the vanishing set of a
non-zero polynomial $\nu\in \QQ[\by][\bx]$. Now note that it suffices
to define $\mathscr{X}_i$ as the complement of the vanishing set of
the coefficients of $\nu$ when it is seen in $\QQ[\bx][\by]$ and
$\mathscr{X} = \cap_{i=1}^{d+1} \mathscr{X}_i$.
\end{proof}
We prove the correctness of Algorithm~\ref{algo:QE} in
Proposition~\ref{prop:correctness} below.
\begin{proposition}
  \label{prop:correctness}
Assume that Assumptions~\eqref{assumption:B2} and \eqref{assumption:A}
hold. Let $\mathscr{O} \subset {\rm GL}(n,t,\CC)$ and $\mathscr{X}
\subset \CC^d$ be defined respectively in
Proposition~\ref{proposition:Noether-parametric} and
Lemma~\ref{lemma:dimension-parametric}. Then for $A \in \mathscr{O}
\cap {\rm GL}(n,t,\QQ)$ and $\alpha \in \mathscr{X}\cap \QQ^d$, the
formula $\Phi$ computed by Algorithm~\ref{algo:QE} defines a dense
subset of the interior of $\pi(\cV_{\RR})$.
\end{proposition}
\begin{proof}
By Lemma~\ref{lemma:dimension-parametric}, $W_i^{A,\alpha}$ satisfies
the assumptions of {\sf RealRootClassification}. Thus, the calls of
{\sf RealRootClassification} on $W_i^{A,\alpha}$ are valid and return
the formulas $\Phi_i$ and the polynomials $w_{i,\infty}$. As $A$ acts
only on $\bx$, $\pi(\cV_{\RR}^{A}) = \pi(\cV_{\RR})$. Thus,
\[Z(\Phi_i) \subset \pi(V(W_i^{A,\alpha})\cap
    \RR^{n+t}) \subset \pi(\cV_{\RR}^{A}) = \pi(\cV_{\RR}).\]
Therefore, $Z(\Phi) = \cup_{i=1}^{d+1} Z(\Phi_i) \subset
\pi(\cV_{\RR})$.

By the description of $\Phi_i$, for $1 \le i \le d+1$,
\[ Z(\Phi_i) \setminus V(w_{i,\infty})= \pi(V(W_i^{A,\alpha})\cap \RR^{n+t})
\setminus V(w_{i,\infty}).\]
Let $\mathscr{Y}_A$ be the non-empty Zariski open subset of $\CC^t$ in
Proposition~\ref{proposition:Noether-parametric} ($\mathscr{Y}_A$
depends on the matrix $A$). We denote 
\[\mathcal{W} = \cup_{i=1}^{d+1} V(w_{i,\infty}) \cup (\CC^t \setminus
  \mathscr{Y}_A).\]
We will show that, for $\eta \in \pi(\cV_{\RR}^{A}) \setminus
\mathcal{W}$, $\eta \in Z(\Phi)$.

Since $\eta \in \pi(\cV_{\RR}^{A})$, $V(\ff^{A}(\eta,\cdot))\cap \RR^n$ is
not empty. On the other hand, as $\eta \in \mathcal{Y}_{A}$,
$\ff^{A}(\eta,\cdot)$ generates a radical equi-dimensional ideal whose
algebraic set is either empty or smooth of dimension $d$. By
Proposition~\ref{prop:sampling}, $V(\ff^{A}(\eta,\cdot))\cap \RR^n$ is
not empty if and only if $\cup_{i=1}^{d+1} V(W_i^{A,\alpha}(\eta) \cap
\RR^n)$ is not empty either. We deduce that $\eta \in \cup_{i=1}^{d+1}
\pi(V(W_{i}^{A,\alpha})\cap \RR^{n+t}) \setminus \mathcal{W}$. We have that
\begin{align*}
& \cup_{i=1}^{d+1} \pi(V(W_{i}^{A,\alpha})\cap \RR^{n+t}) \setminus
\mathcal{W} = \cup_{i=1}^{d+1} (\pi(V(W_{i}^{A,\alpha})\cap \RR^{n+t})
\setminus \mathcal{W}) \\
& = \cup_{i=1}^{d+1} (Z(\Phi_i) \setminus \mathcal{W}) =
(\cup_{i=1}^{d+1} Z(\Phi_i)) \setminus \mathcal{W}.
\end{align*}
Therefore, $Z(\Phi) \setminus \mathcal{W} = \pi(\cV_{\RR}) \setminus
\mathcal{W}$ and $\pi(\cV_{\RR}) \setminus Z(\Phi)$ is of measure zero
in $\RR^t$. By Assumption~\eqref{assumption:A}, we conclude that
$Z(\Phi)$ is a dense subset of the interior of $\pi(\cV_{\RR})$.
\end{proof}
\section{Complexity analysis}
\label{section:complexity}
We now estimate the arithmetic complexity of Algorithm~\ref{algo:QE}
once $A \in \mathscr{O}\cap {\rm GL}(n,t,\QQ)$ and $\alpha \in
\mathscr{X}\cap \QQ^n$ as in Proposition
\ref{proposition:Noether-parametric} are found from a random
choice. In this section, the input $\ff$ forms a regular sequence of
$\QQ[\bx,\by]$ (then, $s=n-d$) satisfying
Assumptions~\eqref{assumption:B2} and \eqref{assumption:A}. As the
calls to {\sf RealRootClassification} on the systems $W_i^{A,\alpha}$
are the most costly parts of our algorithm, we focus on estimating
their complexities. To this end, we introduce the following
assumption.
\begin{assumption}{C}
\label{assumption:Noether}
Let $F \subset \QQ[\bx,\by]$ and $G$ be the reduced Gr\"obner basis of
$F$ w.r.t. the $\DRL(\bx)\succ\DRL(\by)$ order. Then $F$ is said to
satisfy Assumption~\eqref{assumption:Noether} if and only if for any
$g \in G$, the total degree of $g$ in both $\bx$ and $\by$ equals the
degree of $g$ w.r.t. only $\bx$.
\end{assumption}

In \cite[Lemma 13]{LS20}, it is proven that, on an input $F$
satisfying Assumption~\eqref{assumption:Noether}, the polynomial
$w_{\infty}$ in {\sf RealRootClassification} is simply $1$ and the
entries of the Hermite matrix $H_{F}$ are in $\QQ[\by]$. Therefore,
the {\sf SamplePoints} subroutine is called on the sequence of leading
principal minors of the parametric Hermite matrices. Again, with
Assumption~\eqref{assumption:Noether}, the degree of these leading
principal minors can be bounded (see \cite[Lemma
32]{LS20}). Therefore, one obtains the complexity bound for {\sf
  RealRootClassification} for such $F$.

Back to our problem, we will establish a degree bound for the
polynomials given into {\sf SamplePoints}. Some notations that will be
used further are introduced below.

Let $D$ be a bound of the total degree of elements of $\ff$. The
zero-dimensional ideal of $\QQ(\by)[\bx]$ generated by $W_i^{A,
\alpha}$ is denoted by $\langle W_i^{A, \alpha} \rangle$. The quotient
ring $\QQ(\by)[\bx]/ \langle W_i^{A,\alpha} \rangle$ is a finite
dimensional $\QQ(\by)$-vector space. Let $G_i$ be the reduced
Gr\"obner basis of the ideal of $\QQ[\bx,\by]$ generated by $W_i^{A,
\alpha}$ w.r.t. $\DRL(\bx) \succ \DRL(\by)$ and $B_i$ be
the monomial basis of $\QQ(\by)[\bx]/\langle W_i^{A, \alpha} \rangle$
constructed using $G_i$ as in Section \ref{section:preliminary}. We
begin with the following lemma.
\begin{lemma}
  \label{lemma:degree-Noether}
When Assumption~\eqref{assumption:Noether} holds for $W_i^{A,\alpha}$, any
leading principal minor of the matrix $H_i$ has degree bounded
by $2\sum_{b \in B_i} \deg(b)$.
\end{lemma}
\begin{proof}
The proof can be deduced from \cite[Lemma 13, Proposition 31,
Lemma 32]{LS20}. It is mainly based on the control of degrees
appearing in the normal form computation in $\QQ(\by)[\bx]/\langle
W_i^{A,\alpha} \rangle$.
\end{proof}
It remains to estimate the sum $\sum_{b \in B_i}
\deg(b)$. A bound is obtained by simply taking the product of the
highest degree appeared in $B_i$ and its cardinality.  As the
Hilbert series of $\QQ(\by)[\bx]/ \langle W_i^{A,\alpha} \rangle $
when $\ff$ is a generic system are known (see, e.g., \cite{FSSp13,
Spa14}), explicit bounds of these quantities are easily obtained.
\begin{lemma}
  \label{lemma:degree-sum}
Let $B_i$ be defined as above. There exists a dense Zariski open
subset $\mathscr{Q}$ of $\CC[\bx,\by]_{\leq D}^s$ such that, for $\ff
\in \mathscr{Q}$, the following inequality holds for $1\leq i \leq d+1$:
\[\textstyle{\sum_{b \in B_i} \deg_{\bx}(b) \leq (n+s-i)\ D^s(D-1)^{n-i-s+2}\
\binom{n-i+1}{s}}.\]
\end{lemma}
\begin{proof}
By \cite[Theorem 2.2]{NR09}, there exists a dense Zariski open
subset $\mathscr{Q}_{1,1} \subset \CC[\bx,\by]_{\leq D}^s$ such that for
$\ff \in \mathscr{Q}_{1,1}$, the degree of $\langle
W_1^{A,\alpha} \rangle$, which equals to the cardinality of $B_1$, is
bounded by
\[\textstyle{D^s\sum_{k=0}^{n-s} \binom{k+s-1}{s-1}\ (D-1)^k \leq
\ D^s(D-1)^{n-s} \binom{n}{s}}.\]
On the other hand, by \cite[Corollary 3.2]{Spa14}, there exists a
dense Zariski open subset $\mathscr{Q}_{1,2} \subset \CC[\bx,\by]_{\leq
D}^s$ such that for $\ff \in \mathscr{Q}_{1,2}$, the witness degree,
i.e., the highest degree appeared in the reduced Gr\"obner basis of
$W_{1}^{A,\alpha}$ w.r.t. $\DRL(\bx)$, is bounded by
$(n+s-1)D-2n+2$. Thus, the highest degree in $B_{1}$ is bounded by
$(n+s-1)D-2n+1$. Thus, let $\mathscr{Q}_1 = \mathscr{Q}_{1,1} \cap
\mathscr{Q}_{1,2}$ and, for $\ff \in \mathscr{Q}_1$, we obtain
\[\textstyle{\sum_{b\in B_1} \deg(b) \leq (n+s-1)\
D^s(D-1)^{n-s+1}\ \binom{n}{s}}.\]
For $1\leq i\leq d$, the system $W_i^{A,\alpha}$ can also be
interpreted as the system defining the critical locus of the
projection $(x_i,\ldots,x_n) \mapsto x_i$ restricted to $V\left
(\ff^{A}(\alpha_1,\ldots,\alpha_{i-1}, x_{i}, \ldots, x_n) \right
)$. Therefore, by replacing $n$ by $n-i+1$ in the above bound, we
deduce that, for $1\leq i\leq d$, there exists a dense Zariski
open subset $\mathscr{Q}_i \subset \CC[\bx,\by]_{\leq D}^s$ such that
\[\textstyle{\sum_{b\in B_i} \deg(b) \leq (n+s-i)\ D^s(D-1)^{n-i-s+2}\
\binom{n-i+1}{s}}.\]
For $i=d+1$, the cardinality of $B_{d+1}$ is
bounded by $D^s$ and the highest degree in $B_{d+1}$ is bounded by
$s(D-1)$. Thus, the bound holds for $i=d+1$. Taking $\mathscr{Q} =
\cap_{i=1}^{d+1} \mathscr{Q}_i$, we conclude the proof.
\end{proof}
Further, $\mathcal{D}$ denotes $2
(n+s-1)D^s(D-1)^{n-s+1}\binom{n}{s}$. Now we show that
Assumption~\eqref{assumption:Noether} holds generically then prove
Theorem~\ref{thm:main}.
\begin{proposition}
\label{prop:generic-D}
There exists a dense Zariski open subset $\mathscr{P} \subset
\CC[\bx,\by]_{\leq D}^s$ such that, for every $\ff \in \mathscr{P}$,
there exists a dense Zariski open subset $\mathscr{K}_{\ff}
\subset {\rm GL}(n,t,\CC)\times \CC^n$ such that for $(A,\alpha) \in
\mathscr{K}_{\ff}$, Assumption~\eqref{assumption:Noether} holds for every
system $W_i^{A,\alpha}$.
\end{proposition}
\begin{proof}
Let $y_{t+1}$ be a new variable and ${}^h\QQ[\bx,\by,y_{t+1}]_{D}$
be the set of homogeneous polynomials in $\QQ[\bx,\by,y_{t+1}]$ of
degree $D$. For $F \subset \QQ[\bx,\by]$, we denote by
${}^hF \subset \QQ[\bx,\by,y_{t+1}]$ the homogenization of $F$
w.r.t. all the variables $(\bx,\by)$, that means ${}^hF =
y_{t+1}^{\deg(p)}\cdot F \left( \frac{x_1}{y_{t+1}}, \ldots,
\frac{x_n}{y_{t+1}}, \frac{y_1}{y_{t+1}}, \ldots,
\frac{y_t}{y_{t+1}}\right)$ for each $p \in F$. Further, $\langle
{}^hF \rangle_h$ denotes the ideal of $\CC[\bx,\by,y_{t+1}]$ generated
by ${}^hF$.

We consider the following property {\sf (C1)}: The leading terms
appearing in the reduced Gr\"obner basis of $\langle {}^hF \rangle_h$
w.r.t. $\DRL(\bx \succ \by \succ y_{t+1})$ do not involve
any of the variables $y_1,\ldots,y_{t+1}$. By the proof of
\cite[Prop. 30]{LS20}, the property \textsf{(C1)}
implies Assumption \eqref{assumption:Noether}.

Following the proof of \cite[Prop. 7]{BFS14}, if $y_{j+1}$ is not a
zero-divisor of the quotient ring $\CC[\bx,\by,y_{t+1}]/\langle {}^hF,
y_{1},\ldots,y_{j} \rangle_h$ for every $0 \leq j \leq t$, then $F$
satisfies the property \textsf{(C1)}. This property means that
$(y_1,\ldots,y_{t+1})$ forms a regular sequence in the quotient ring
$\CC[\bx,\by,y_{t+1}]/ \langle {}^h F \rangle_h$. We name this
property as \textsf{(C2)}. 

From the proof of \cite[Lemma 2.1, Lemma 2.2]{Spa14} and
\cite[Proposition 18.13]{Eisenbud}, there exists a dense Zariski open
subset $\mathscr{P}_1 \subset \CC[\bx,\by]_{\leq D}^s$ such that for
$\ff \in \mathscr{P}_1$, there exists a dense Zariski open subset
$\mathscr{K}_{\ff,1} \subset {\rm GL}(n,t,\CC)\times \CC^n$ such that
for $(A,\alpha) \in \mathscr{K}_{\ff,1}$, the quotient ring
$\CC[\bx,\by,y_{t+1}]/ \langle {}^hW_1^{A,\alpha} \rangle_h$ is a
Cohen-Macaulay ring of dimension $t+1$ and the ideal $\langle
{}^hW_1^{A,\alpha}, y_1,\ldots,y_{t+1} \rangle_h$ has dimension
$0$. By the unmixedness theorem \cite[Corollary 18.14]{Eisenbud},
$(y_1,\ldots,y_{t+1})$ is a regular sequence over
$\CC[\bx,\by,y_{t+1}]/\langle {}^hW_1^{A,\alpha} \rangle_h$. Thus,
$W_1^{A,\alpha}$ satisfies the property \textsf{(C2)} and
Assumption~\eqref{assumption:Noether} holds.

Similar for $2 \leq i \leq d+1$, we obtain dense Zariski subsets
$\mathscr{P}_i \subset \CC[\bx,\by]_{\leq D}^s$ and
$\mathscr{K}_{\ff,i} \subset {\rm GL}(n,t,\CC)\times \CC^n$ for each $\ff
\in \mathscr{P}_i$. Taking $\mathscr{P} = \cap_{i=1}^{d+1}
\mathscr{P}_i$, and $\mathscr{K}_{\ff} = \cap_{i=1}^{d+1}
\mathscr{K}_{\ff,i}$, we conclude the proof.
\end{proof}
\renewcommand*{\proofname}{Proof of Theorem~\ref{thm:main}}
\begin{proof}
It is well-known that Assumptions~\eqref{assumption:B2}
and \eqref{assumption:A} are generic. Also, the set of regular
sequences is dense in $\CC[\bx,\by]_{\leq D}^s$. Thus, there
exists a dense Zariski open subset $\mathscr{R} \subset
\CC[\bx,\by]_{\leq D}^s$ such that for any $\ff \in \mathscr{R}$,
$\ff$ forms a regular sequence satisfying
Assumptions~\eqref{assumption:B2} and \eqref{assumption:A}. As
$V(\ff)$ has dimension $d+t$ and $\ff$ forms a regular sequence in
$\QQ[\bx,\by]$, $d = n -s$. Algorithm~\ref{algo:QE} consists of
$(d+1)$ calls to {\sf RealRootClassification} on $W_i^{A,\alpha}$. Let
$\mathscr{P}$ be the dense Zariski open set in
Proposition~\ref{prop:generic-D} and $\mathscr{Q} = \mathscr{P} \cap
\mathscr{R}$. Then, for $\ff \in \mathscr{Q}$, {\sf SamplePoints} is
called on a list of polynomials in $\QQ[\by]$ of degree bounded by
$\mathcal{D}$. The number of principal minors is equal to the
dimension of the quotient ring $\QQ(\by)[\bx]/\langle W_i^{A,\alpha}
\rangle $, which is also bounded by $\mathcal{D}$. Applying
\cite[Theorem 2]{LS20}, each call to {\sf RealRootClassification} on
$W_i^{A,\alpha}$ costs at most $O\ {\widetilde{~}} \left( 8^{t}\
\mathcal{D}^{3t+2}\ \binom{t+\mathcal{D}}{t} \right)$ arithmetic
operations in $\QQ$. In total, the arithmetic complexity of
Algorithm~\ref{algo:QE} is bounded by $O\ {\widetilde{~}}\left
((n-s+1)\ 8^{t}\ \mathcal{D}^{3t+2}\ \binom{t+\mathcal{D}}{t}\right
)$.
\end{proof}

\section{Experiments}
\label{section:experiment}
\renewcommand{\figurename}{Table}
We compare the practical behavior of Algorithm \ref{algo:QE} with
QuantifierElimination (\textsc{maple}'s RegularChains) and Resolve
(\textsc{mathematica}) on an Intel(R) Xeon(R) Gold 6244 3.60GHz
machine of 754GB RAM. The timings are given in seconds (s.), minutes
(m.) and hours (h.). The symbol $\infty$ means that the computation is
stopped after 72 hours without getting the result. We use our
\textsc{maple} implementation for Hermite matrices, in which
\textsc{FGb} package \cite{FGb} is used for Gr\"obner bases
computation. The computation of sample points is done by
\textsc{RAGlib} \cite{RAG} which uses \texttt{msolve} \cite{MSolve}
for polynomial system solving.

For {\sf RealRootClassification}, we use the following
notations:
\begin{itemize}
\item {\sc hm}: timings of computing Hermite matrices and their
minors.
\item {\sc sp}: total timings of computing the sample points.
\item {\sc size}: the largest size of the Hermite matrices.
\item {\sc deg}: the highest degree appeared in the output formulas.
\end{itemize}

Start with random dense systems, we fix the total degree $D = 2$ and
run our algorithm for various $(t,n,s)$. In
Table~\ref{fig:D-2-generic}, {\sf SamplePoints} accounts for the major
part of our timings. While our
algorithm can tackle these examples, neither {\sc maple} nor {\sc
mathematica} finish within 72h. The theoretical degree bound agrees
with the practical observations. This agrees with our complexity
result. On smaller problems, we observe that formulas computed by {\sc
maple} and {\sc mathematica} have larger degrees than our
output. Hence, these implementations, based on CAD, suffer from its
doubly exponential complexity while our implementation takes advantage
of the singly exponential complexity of our algorithm.
\begin{figure}[!ht]
  \footnotesize
  \centering
  \begin{tabular}{|c c c | c c | c | c | c | c |}
    \toprule
  $t$ & $n$ & $s$ & {\sc hm} & {\sc sp} & {\sc size} & {\sc deg} &
{\sc maple} & {\sc mathematica} \\
  \midrule
  $2$ & $3$ & $2$ & .2 s.& 3 s. & 8 & 24 & $\infty$ & $\infty$ \\
  $2$ & $4$ & $2$ & 9 s.& 1 m. & 12 & 40 & $\infty$ & $\infty$\\
  $2$ & $5$ & $2$ & 2 m.& 15 m. & 16 & 56 & $\infty$ & $\infty$ \\
  $2$ & $6$ & $2$ & 20 m.& 2.5 h. & 20 & 72 & $\infty$ & $\infty$ \\
  $2$ & $7$ & $2$ & 1.5 h. & 6 h. & 24 & 88 & $\infty$ & $\infty$
    \\  
  \midrule
  $3$ & $3$ & $2$ & 6 s. & 1 m. & 8 & 24 & $\infty$ & $\infty$ \\
  $3$ & $4$ & $2$ & 5 m. & 15 m. & 12 & 40 & $\infty$ & $\infty$ \\
  $3$ & $5$ & $2$ & 2 h. & 5 h. & 16 & 56 & $\infty$ & $\infty$ \\
  $3$ & $6$ & $2$ & 8 h. & 16 h. & 20 & 72 & $\infty$ & $\infty$ \\
  \midrule
  $4$ & $3$ & $2$ & 40 s. & 30 m. & 8 & 24 & $\infty$ & $\infty$ \\
  $4$ & $4$ & $2$ & 6 h. & 40 h. & 12 & 40 & $\infty$ & $\infty$ \\
  \midrule
  $5$ & $3$ & $2$ & 5 m. & 14 h. & 8 & 24 & $\infty$ & $\infty$ \\
  \bottomrule
\end{tabular}
\caption{Generic systems with $D=2$}
\label{fig:D-2-generic}
\end{figure}

Table~\ref{fig:sparse} shows the timings for sparse systems. Each
polynomial is generated with $D=2$ and has $2n$ terms. Even
Assumption~\eqref{assumption:Noether} is not satisfied, our algorithm
still applies. Thanks to the sparsity, the size and degree of the
matrices in our algorithm are smaller than in the dense cases. Thus,
our algorithm runs faster here than in Table~\ref{fig:D-2-generic}
while these examples are out of reach of \textsc{maple} and
\textsc{mathematica}.
\begin{figure}[!ht]
  \footnotesize
  \centering
  \begin{tabular}{|c c c | c c | c | c | c | c |}
  \toprule
  $t$ & $n$ & $s$ & {\sc hm} & {\sc sp} & {\sc size} & {\sc deg} &
{\sc maple} & {\sc mathematica} \\
  \midrule
  $3$ & $3$ & $2$ & 3 s. & 37 s. & 7 & 22 & $\infty$ & $\infty$ \\
  $3$ & $4$ & $2$ & 2 m. & 10 m.  & 9 & 34 & $\infty$ & $\infty$ \\
  $3$ & $5$ & $2$ & 2 m. & 10 m. & 9 & 32 & $\infty$ & $\infty$ \\
  \midrule 
  $4$ & $3$ & $2$ & 20 s. & 20 m. & 7 & 22 & $\infty$ & $\infty$ \\
  $4$ & $4$ & $2$ & 15 s. & 18 m. & 5 & 20 & $\infty$ & $\infty$ \\
  \bottomrule
\end{tabular}
\caption{Sparse systems with $D=2$}
\label{fig:sparse}
\end{figure}

Table~\ref{fig:structure} gives the timings for structured
systems. We separate the variables $\bx$ into blocks of total degree
$1$; $[i,n-i]$ means that the degree in $[x_1,\ldots,x_i]$ and
$[x_{i+1},\ldots,x_n]$ are respectively $1$. Here, entries of the
Hermite matrices have non-trivial denominators with high
degree. Computation those matrices takes the major part. However, our
algorithm still outperforms the two other software.
\begin{figure}[!ht]
  \footnotesize
  \centering
  \begin{tabular}{|c c c c | c c | c | c | c | c |}
    \toprule
  $t$ & $n$ & $s$ & Block & {\sc hm} & {\sc sp} & {\sc size} & {\sc
deg} & {\sc maple} & {\sc mathematica} \\
  \midrule
    $3$ & $3$ & $2$ & $[1,2]$ & 5 s. & 45 s. & 4 & 20 & $\infty$ &
$\infty$ \\
    $3$ & $4$ & $2$ & $[2,2]$ & 4 m. & 1 m. & 8 & 32 & $\infty$ &
$\infty$\\
    $3$ & $5$ & $2$ & $[2,3]$ & 2 h. & 9 m. & 8 & 40 & $\infty$ &
$\infty$\\
    $3$ & $6$ & $2$ & $[3,3]$ & 30 h. & 45 m. & 14 & 60 & $\infty$ &
$\infty$\\
  \bottomrule
\end{tabular}
\caption{Structured systems}
\label{fig:structure}
\end{figure}

\bibliographystyle{plain}

\bibliography{QEbib}

\end{document}